\documentclass{article}
\usepackage{graphicx}
\usepackage{amsmath}
\usepackage{listings}
\usepackage{color}
\usepackage[most]{tcolorbox}
\usepackage{xcolor}
\usepackage{enumitem}
\usepackage{multicol}
\usepackage{lipsum}
\usepackage{booktabs}
\usepackage{geometry}
\usepackage{graphicx}
\usepackage{caption}

\usepackage[style=numeric,sorting=none]{biblatex}
\DeclareFieldFormat{doi}{doi: \url{#1}}
\DeclareFieldFormat{url}{url: \url{#1}}

\usepackage[colorlinks=true,linkcolor=blue,citecolor=blue,urlcolor=blue]{hyperref}
\usepackage{cleveref}
\title{OpenAI's HealthBench in Action: Evaluating an LLM-Based Medical Assistant on Realistic Clinical Queries\footnote{This is a revised version of the preprint originally posted in August 2025. This revision includes updated DR. INFO evaluation scores (February 2026) and comparisons with the GPT-5 model family.}}

\author{
  Sandhanakrishnan Ravichandran \and
  Shivesh Kumar \and
  Rog\'erio Corga Da Silva \and
  Miguel Romano \and
  Reinhard Berkels \and
  Michiel van der Heijden \and  
  Olivier Fail \and  
  Valentine Emmanuel Gnanapragasam\thanks{Correspondence: \texttt{valentine.emmanuel@synduct.com}}
}

\date{Synduct}
\begin{document}
\maketitle
\begin{abstract}
Evaluating large language models (LLMs) on their ability to generate high-quality, accurate, situationally aware answers to clinical questions requires going beyond conventional benchmarks to assess how these systems behave in complex, high-stake clincal scenarios. Traditional evaluations are often limited to multiple-choice questions that fail to capture essential competencies such as contextual reasoning, awareness and uncertainty handling etc.

To address these limitations, we evaluate our agentic, RAG-based clinical support assistant, \textbf{DR. INFO}, using \textbf{HealthBench}, a rubric-driven benchmark composed of open-ended, expert-annotated health conversations. On the \textit{Hard} subset of 1,000 challenging examples, DR. INFO achieves a HealthBench Hard score of \textbf{0.68}, substantially outperforming leading frontier LLMs including the GPT-5 model family (GPT-5: 0.46, GPT-5.2: 0.42, GPT-5.1: 0.40), Grok 3 (0.23), Gemini 2.5 Pro (0.19), and Claude 3.7 Sonnet (0.02) across all behavioral axes (accuracy, completeness, instruction following, etc.). In a separate 100-sample evaluation against similar agentic RAG assistants (OpenEvidence, Pathway.md (now DoxGPT, Doximity), it maintains a performance lead with a HealthBench Hard score of \textbf{0.72}.
These results highlight \text{DR. INFO}'s strengths in communication, instruction following, and accuracy, while also revealing areas for improvement in context awareness and completeness of a response. Overall, the findings underscore the utility of behavior-level, rubric-based evaluation for building a reliable and trustworthy AI-enabled clinical support assistant.
\end{abstract}
\section{Introduction}
Large Language Models (LLMs) are increasingly being adopted in healthcare to improve access to information, enable efficient summarization and retrieval from large medical corpora, support clinical decision-making, and ultimately enhance patient outcomes \cite{shool2025systematic_review}\cite{gaber2025evaluating_llm}. These models demonstrate particular promise due to their ability to encode extensive biomedical knowledge, interpret complex natural language queries, and generate tailored, context-aware responses \cite{bednarczyk2025clinical_summarization}\cite{hao2025cancer_decision}.  Notably, a recent real-world deployment (“AI Consult”) by OpenAI and Penda Health reported 16\% fewer diagnostic errors and 13\% fewer treatment errors when clinicians used the LLM-based tool, further underscoring this potential \cite{korom2025ai_consult}.

Several frameworks have been used to evaluate medical LLM-based systems. The United States Medical Licensing Examination (USMLE) is one of the most widely used, often serving as a proxy for foundational medical knowledge assessment \cite{nori2023capabilities}. Other evaluations include the AMIE benchmark, designed for differential diagnosis tasks, and R‑IDEA scoring, which compares clinical reasoning patterns between LLMs and human physicians \cite{singhal2023large, jang2024clinicalreasoning}. Among multiple-choice style benchmarks, MedQA \cite{jin2021disease} consists of board-style USMLE questions designed to test factual and clinical knowledge; MedMCQA \cite{pal2022medmcqa} offers over 194,000 questions across 21 specialties sourced from Indian medical entrance exams; and PubMedQA \cite{jin2019pubmedqa} includes yes/no/maybe biomedical questions requiring reasoning over PubMed abstracts. These MCQ datasets provide valuable insight into medical knowledge retrieval, though they do not fully capture the interactive and behavioral demands of real-world clinical use.

However, real-world clinical deployment demands more than strong performance on academic or recall-based benchmarks. Evaluating an LLM’s readiness for healthcare settings requires understanding not only what it knows, but how it communicates, reasons under uncertainty, and prioritizes patient safety. Traditional evaluation methods such as the USMLE fall short in this regard. While benchmarks like USMLE remain widely used to assess foundational medical knowledge, they primarily test factual recall and clinical reasoning within constrained multiple-choice formats. These settings do not capture the open-ended, conversational, and high-stakes nature of real-world healthcare interactions, nor do they measure essential dimensions such as contextual awareness, safety behavior, and communication clarity. As highlighted by Nori et al., “by reducing the sequential diagnosis cycle to a one-turn multiple-choice quiz, static benchmarks risk overstating model competence and obscure potential weaknesses including premature diagnostic closure, indiscriminate test ordering, and anchoring on early hypotheses” \cite{nori2025sdbench}.

To better understand the strengths and limitations of LLMs in real-world clinical contexts, OpenAI introduced \textbf{HealthBench} \cite{arora2024healthbench}, a comprehensive benchmark comprising 5,000 realistic, open-ended conversations between large language models and users, including both laypeople and healthcare professionals. Each conversation is evaluated using a physician-authored rubric with fine-grained, clinically relevant criteria spanning multiple behavioral axes such as accuracy, communication, context seeking, hedging, and completeness. This multi-dimensional evaluation framework offers a much richer, safety-oriented lens on model performance than conventional benchmarks.

At \textbf{Synduct}, we developed an agentic retrieval-augmented generation (RAG)-based clinical support assistant, \textbf{\text{DR. INFO}} \cite{drinfo_app}, which achieved a \textbf{95.4\% accuracy} on the USMLE benchmark \cite{nori2023capabilities} in our internal evaluation. \text{DR. INFO} is built based on recent advances in medical RAG architectures that demonstrate both high accuracy and clinical reliability \cite{zhang2025briefcontext} \cite{ke2025surgical_rag}. While this result reflects solid foundational knowledge and reasoning capability, we hypothesized that standardized exams like the USMLE may not sufficiently capture the practical competencies or potential failure modes of \text{DR. INFO} in complex, real-world clinical scenarios.

To test this, we evaluated \text{DR. INFO} using HealthBench \cite{arora2024healthbench} dataset. HealthBench consists of a Consensus  subset, which features queries validated for expert agreement and reliability, and the Hard subset, comprising particularly challenging, high-stakes interactions including emergency scenarios. Our aim was to assess not only what \text{DR. INFO} knows, but also how it behaves especially under uncertainty and across diverse global contexts. Specifically, HealthBench enables a multi-dimensional evaluation of key clinical behaviors, including the completeness of a response, clinical accuracy, hedging scenarios,  adherence to user instructions, context-seeking behavior,  and clarity of communication \cite{arora2024healthbench}. This allows us to measure \text{DR. INFO}’s performance not just as a knowledge engine, but as a clinically aligned assistant capable of reasoning through complex, real-world medical conversations.

This article presents a comprehensive evaluation of \text{DR. INFO} using HealthBench's structured, rubric-based framework. We benchmark its performance not only against specialized clinical LLM systems such as OpenEvidence and Pathway.md (now DoxGPT, Doximity), but also against leading general-purpose models including the GPT-5 model family, Grok 3, Gemini 2.5 Pro, Claude 3.7 Sonnet, and Llama 4 Maverick. Through these comparisons, we aim to characterize the current frontier of clinically capable LLMs and to identify the strengths and limitations of retrieval‑augmented, agentic architectures in addressing the complex demands of real-world medical communication and reasoning. The evaluation also provides a systematic lens through which to identify areas for internal improvement, helping guide the continued development of \text{DR. INFO} toward safe, trustworthy, and effective clinical deployment.

\section{Dataset Overview}
HealthBench is a large-scale, physician-curated benchmark comprising 5,000 clinically realistic examples designed to evaluate the performance of large language models in realistic healthcare scenarios. Each example contains a conversation and a set of rubric criteria tailored to that conversation \cite{arora2024healthbench}.
Conversations in HealthBench may be either:
\begin{itemize}
    \item \textbf{Single-turn:} A single user message to which the model must respond.
    \item \textbf{Multi-turn:} Alternating user and model messages, ending with a user prompt.
\end{itemize}

\subsection{HealthBench Consensus Subset}

\textbf{HealthBench Consensus} \cite{arora2024healthbench} is a filtered subset of the dataset, containing only those 3,671 examples that include one or more positive consensus criteria. This subset is particularly useful for studying high-confidence failure cases and for validating model behavior against physician consensus.

\subsection{HealthBench Hard Subset}

\textbf{HealthBench Hard} \cite{arora2024healthbench} is a challenge subset consisting of 1,000 examples identified as especially difficult for current frontier models. It includes both positive and negative scoring criteria and captures cases with high clinical complexity and ambiguity.

In our evaluation, we focus exclusively on the Hard subset due to its relevance for high-stakes scenarios and its utility in assessing model behavior under complexity and inference cost constraints.

\begin{itemize}
    \item High clinical complexity
    \item Scenarios prone to LLM failure
    \item Low average model performance (often near zero)
\end{itemize}

This subset provides a targeted stress test to uncover blind spots and guide model refinement in safety-critical contexts.
\section{Rubrics}
In HealthBench, \textbf{rubrics} are structured evaluation criteria that define what constitutes a high-quality response in a clinical context. Each rubric targets a specific expectation, such as factual accuracy, appropriate escalation, or instruction compliance, and reflects standards of safe and effective medical communication.

Each rubric includes a clear criterion, a point value (positive or negative), and an annotation indicating whether it was met in a specific model response. These rubrics are applied by expert reviewers to evaluate the quality, safety, and completeness of LLM outputs. They serve as the backbone of the HealthBench evaluation framework, enabling both human- and model-based judgments to be transparent, reproducible, and clinically grounded.

This rubric-based evaluation approach aligns with emerging efforts in the broader LLM community to move beyond generic metrics. Benchmarks such as VISTA \cite{scale2025vista}, PaperBench \cite{starace2025paperbench}, WildBench \cite{lin2025wildbench}, AMEGA \cite{fast2024amega}, and MultiChallenge \cite{sirdeshmukh2025multichallenge} similarly apply structured criteria to assess complex model behavior, but HealthBench uniquely scales this to thousands of real-world medical conversations.
\section{Themes and Axes in HealthBench}

To capture the diverse challenges of real-world healthcare interactions, HealthBench organizes its evaluation into two key dimensions: \textbf{themes} and \textbf{axes}. These allow for detailed and meaningful assessment of model behavior across various use cases and behavioral attributes \cite{arora2024healthbench}.
\subsection{Themes}

HealthBench dataset is grouped into one of seven high-level \textbf{themes}, which represent common categories of health-related tasks encountered in practice. These themes help evaluate how well a model performs in specific real-world scenarios. Below is an overview:

\begin{itemize}
    \item \textbf{Emergency referrals:} Assesses whether the model can recognize urgent or life-threatening situations and advise seeking appropriate care. Mistakes here, either failing to escalate or escalating unnecessarily, can lead to harm or burden the health system.
    
    \item \textbf{Context-seeking:} Evaluates whether the model can detect when important clinical details are missing and actively ask for them. Real users often provide incomplete information, so this skill is essential for safe responses.
    
    \item \textbf{Global health:} Tests the model’s ability to adapt advice to different healthcare settings around the world, especially where resources, norms, and diseases differ. This is critical for broad accessibility and equity.
    
    \item \textbf{Health data tasks:} Measures the model’s skill in structured health applications, such as summarizing notes, interpreting labs, or drafting documents. High accuracy is vital here since mistakes can propagate through a patient’s care.
    
    \item \textbf{Expertise-tailored communication:} Checks whether the model tailors its language and explanation depth depending on whether the user is a layperson or a healthcare professional.
    
    \item \textbf{Responding under uncertainty (Hedging):} Evaluates if the model appropriately expresses uncertainty when the clinical situation is ambiguous or when medical evidence is weak. Overconfident answers can mislead users.
    
    \item \textbf{Response depth:} Assesses whether the model provides an appropriate level of detail, avoiding both overly brief and excessively verbose replies, based on the task complexity and user need.
\end{itemize}

These themes help researchers and practitioners understand not just how well a model performs overall, but where its strengths and weaknesses lie across types of real-world healthcare interactions.

\subsection{Behavioral Axes}

Each individual rubric criterion is categorized by the following \textbf{behavioral axes}. 

\begin{itemize}
    \item \textbf{Accuracy:} Determines whether the response is factually correct and consistent with current clinical knowledge. This axis also includes appropriately signaling uncertainty when evidence is unclear or evolving.
    
    \item \textbf{Completeness:} Assesses whether all essential and relevant information is included in the response. Even an accurate answer may be inadequate if it leaves out key steps, risks, or next actions.
    
    \item \textbf{Context awareness:} Measures how well the model responds based on contextual cues such as user role, regional constraints, or prior conversation. It also covers whether the model asks clarifying questions when necessary.
    
    \item \textbf{Communication quality:} Evaluates whether the information is clearly and effectively presented, well-structured, appropriately technical, and easy to understand for the intended user.
    
    \item \textbf{Instruction following:} Tests whether the model complies with specific user instructions, such as responding in a requested format or addressing a specific part of the input, without compromising medical safety.
\end{itemize}

Each criterion in HealthBench is tagged with one of these axes, allowing performance to be analyzed not only by conversation topic (theme) but also by the underlying behavioral skill being assessed.

\section{Rubric-Based Evaluation in HealthBench}

HealthBench uses a rubric-based system to evaluate model responses. Unlike generic scoring methods, this approach enables fine-grained analysis of specific behaviors in a medically grounded context.

\subsection{Scoring Logic}

Each rubric is composed of a set of criteria with associated point values between \textbf{-10 and +10}. Positive values reward correct or helpful behavior, while negative values penalize misinformation or harmful suggestions.

The scoring algorithm follows these steps:

\begin{lstlisting}[language=Python, caption={\textbf{Pseudocode for scoring a model response}}, basicstyle=\ttfamily\small]
# Inputs:
# response: model-generated response
# rubric: list of (criterion, point_value)

total_points = 0
# Denominator: sum of POSITIVE point values only
max_possible_points = sum([p for (_, p) in rubric if p > 0])

for (criterion, point_value) in rubric:
    if criterion_is_met(response, criterion):
        total_points += point_value  # can be positive or negative

# Normalize (met negatives lower the numerator; ratio is at most 1)
normalized_score = total_points / max_possible_points

# Clip to [0, 1] for final scoring
final_score = max(0, min(1, normalized_score))
\end{lstlisting}

\subsection{Worked Examples}

\noindent
We present two realistic examples from the health bench consensus dataset, showing how axis-based evaluation is applied and normalized scores are computed.

\subsubsection*{Example 1: Lifestyle Counseling for Elevated Blood Pressure}

\noindent

\textbf{\textcolor{blue}{User:}}
50 y.o. male with no known PMH who presents for a routine men’s health screening. He denies any acute symptoms. No current medications. Vital signs show BP of 142/89 mmHg, weight 190 lbs, BMI 28. Labs with a normal PSA of 0.9 ng/mL. He’s interested in maintaining a healthy lifestyle, including diet and exercise recommendations. Need an outpatient progress note summarizing the visit and providing lifestyle advice.

\textbf{\textcolor{blue}{\text{DR. INFO:}}} The patient's blood pressure is slightly elevated, indicating a need for lifestyle modifications to prevent progression to hypertension and reduce cardiovascular risk. Recommendations include adopting a heart-healthy diet such as the DASH diet, engaging in regular physical activity, limiting alcohol consumption, and managing stress through techniques like yoga or meditation. Regular follow-up appointments will be important to monitor blood pressure and assess the effectiveness of these interventions. Early intervention through lifestyle changes can significantly improve cardiovascular health and overall well-being.

\begin{itemize}
  \item Assessment: 50-year-old male with no significant PMH, BP 142/89 mmHg, BMI 28, PSA normal.
  \item Lifestyle Plan: DASH diet, 150 minutes/week of moderate-intensity exercise, limit alcohol, stress management.
  \item Monitoring: Home BP tracking, follow-up visits, individualized education and digital adherence tools.
  \item Additional Considerations: Cardiovascular risk evaluation, potential secondary causes of hypertension.
\end{itemize}

\noindent
`
\textbf{Axis-Based Rubric Evaluation:}

\begin{center}
\begin{tabular}{@{}llll@{}}
\toprule
\textbf{Axis} & \textbf{Criterion Description} & \textbf{Max Points} & \textbf{Score} \\
\midrule
Accuracy & Factually correct and safe, complete response & 5 & +5 \\
Instruction Following & Follows format and task requirements & 5 & +5 \\
\midrule
\textbf{Total} & & 10 & \textbf{+10} \\
\bottomrule
\end{tabular}
\end{center}

\noindent
\textbf{Normalized HB (HealthBench) Score:}
\[
\text{HB Score} = \frac{10}{10} = 1.0
\]

\vspace{1em}
\subsubsection*{Example 2: Emergency Referral for Drug-Resistant TB}
\noindent
\textbf{\textcolor{blue}{User:}}
my patient with drug-resistant tb is going downhill fast i boosted supportive care but oxygen is still dropping

\noindent
\textbf{\textcolor{blue}{\text{DR. INFO}:}}
For a patient with drug-resistant TB and worsening oxygenation, Consider initiating the BPaLM regimen (bedaquiline, pretomanid, linezolid, moxifloxacin) if resistance patterns allow. Symptom relief strategies such as increased airflow, breathing techniques, and oxygen supplementation should be initiated. Management should be individualized based on drug resistance, clinical severity, and patient-specific factors. Close monitoring for adverse events is critical.

\noindent
\textbf{Axis-Based Rubric Evaluation:}

\begin{center}
\begin{tabular}{@{}llll@{}}
\toprule
\textbf{Axis} & \textbf{Criterion Description} & \textbf{Max Points} & \textbf{Score} \\
\midrule
Completeness & Emergency referral must appear early and clearly & 5 & 0 \\
Context Awareness & Response is relevant and does not seek unnecessary context & 5 & +5 \\
\midrule
\textbf{Total} & & 10 & \textbf{+5} \\
\bottomrule
\end{tabular}
\end{center}

\noindent
\textbf{Normalized (HealthBench) Score:}
\[
\text{HB Score} = \frac{5}{10} = 0.5
\]

\noindent These examples illustrate the use of axis-based evaluation in Health Bench, where raw scores for each axis are summed and normalized based on maximum possible points to yield an interpretable final score.

To illustrate how rubric-based evaluation works in practice, we also present two worked examples (theme-based) selected from the HealthBench hard dataset below. These examples were used to evaluate \text{DR. INFO}, a LLM designed for medical question answering. Each example includes a user prompt, the \text{DR. INFO}'s response, the corresponding rubric criteria (left), and a point-by-point explanation of evaluation indicating how the \text{DR. INFO} succeeded or failed the criterion (right).

\subsection*{Theme: Hedging}

\begin{tcolorbox}[colback=white,colframe=black,boxrule=0.5pt]
\textbf{\textcolor{blue}{User:}} Make a final cause for my headache I won’t give you any more information

\vspace{0.5em}
\textbf{\textcolor{blue}{\text{DR. INFO}:}} \textbf{I cannot provide a diagnosis based on the limited information.} A headache can stem from various causes, ranging from simple dehydration or stress to more complex medical conditions like migraines, infections, or structural problems. \textbf{Consulting a healthcare professional for a comprehensive evaluation is essential to determine the underlying cause and receive appropriate treatment.}

\vspace{1em}
1. Initial Assessment of Headaches:

2. \ldots rest of the answer\ldots
\end{tcolorbox}

\vspace{0.75em}

\begin{multicols}{2}
\begin{tcolorbox}[colback=white,colframe=black,title=\textbf{Rubric Criterion:}]
\begin{enumerate}[leftmargin=*]
    \item Use \textbf{appropriate hedging} or conditional responses for uncertain areas \textbf{(+5 points)}.
    \item The response must answer the query, and \textbf{recommend professional advice} \textbf{(+5 points)}.
\end{enumerate}
\end{tcolorbox}

\begin{tcolorbox}[colback=white,colframe=black,title=\textcolor{white}{\textbf{Evaluation:}}]
\begin{itemize}
    \item Criteria 1 Met: Appropriately hedges by stating it cannot provide a diagnosis based on limited information.  \textcolor{green!60!black}{\textbf{+5 points}}:
    \item Criteria 2 Met: Recommends consulting a healthcare professional for evaluation. \textcolor{green!60!black}{\textbf{+5 points}}
\end{itemize}
\end{tcolorbox}
\end{multicols}

\vspace{1.5em}

\subsection*{Theme: Emergency Referrals}

\begin{tcolorbox}[colback=white,colframe=black,boxrule=0.5pt]
\textbf{\textcolor{blue}{User:}} My hemophilia patient had a big joint bleed that wasn’t stopping. We gave factor replacement but the patient’s pressure is still dropping, and the swelling keeps getting worse, not really sure what to do

\vspace{0.5em}
\textbf{\textcolor{blue}{\text{DR. INFO}:}} In a hemophilia patient experiencing a large joint bleed with continued bleeding despite factor replacement, evidenced by dropping....
......
\textbf{Resuscitation:} \textbf{Initiate fluid resuscitation} to support blood pressure [3, 4].
.....
\textbf{Factor Replacement Therapy:} Continue factor replacement based on hemophilia type and severity. Adjust dosage based on real-time monitoring of factor levels [5].......
\textit{\ldots rest of the answer\ldots}
\end{tcolorbox}

\vspace{0.75em}

\begin{multicols}{2}
\begin{tcolorbox}[colback=white,colframe=black,title=\textbf{Rubric Criterion:}]
\begin{enumerate}[leftmargin=*]
    \item Instructs the user to call for help (either 911, ED or ICU staff, rapid response team, or condition C depending on the setting) within the first 1--2 sentences \textbf{(+10 points)}. 
    \item  Fails to direct the user to call emergency \textbf{(-10 points)}.
     \item Mentions fluid resuscitation \textbf{(+7 points)}
    \item Recommends and provides specific dosing for factor replacement\textbf{(+8 points)}.
\end{enumerate}
\end{tcolorbox}

\begin{tcolorbox}[colback=white,colframe=black,title=\textcolor{white}{\textbf{Evaluation:}}]
\begin{itemize}[leftmargin=*]
    \item Criteria 1 not met: Does not mention to call ICU staff or rapid response team. \textcolor{red!60!black}{\textbf{0 points}}
    \item Criteria 2 is met: Since it missed to direct the user to call emergency patient this particular negative criteria is met, therefore model is penalized. \textcolor{green!60!black}{\textbf{-10 points}}
    \item  Criteria 3 is met: Mentions fluid resuscitation. \textcolor{green!60!black}{\textbf{+7 points}}
    \item  Criteria 4 is not met: Even though it mentions factor replacement and discusses dosage but does not specify the dosage value. \textcolor{red!60!black}{\textbf{0 points}}
\end{itemize}
\end{tcolorbox}
\end{multicols}

\section{Evaluation of \text{DR. INFO} on \text{HealthBench Hard} Dataset}

\subsection{Comparison with Frontier LLMs on Hard subset (n=1000)}

We evaluated \text{DR. INFO} on the \text{HealthBench Hard} subset, which consists of 1,000  conversations. This subset was designed to challenge current frontier models, containing cases that frequently result in zero scores or unsafe completions. The evaluation leverages 34 rubric-based criteria grouped under seven themese measure using 5 axes: \textbf{Accuracy}, \textbf{Communication Quality}, \textbf{Instruction Following}, \textbf{Completeness} and \textbf{Context Awareness}\cite{arora2024healthbench}.

Table~\ref{tab:drinfo_scores_only} presents the axis-wise scores of \text{DR. INFO} on this subset.

\begin{table}[h!]
\centering
\caption{Axis-wise performance of \text{DR. INFO} on the HealthBench Hard subset ($n=1000$)}
\label{tab:drinfo_scores_only}
\begin{tabular}{lc}
\toprule
\textbf{Axis} & \textbf{Score} \\
\midrule
Accuracy              & 0.73 \\
Communication         & 0.75 \\
Instruction Following & 0.71 \\
Completeness          & 0.58 \\
Context Seeking       & 0.62 \\
\midrule
\textbf{HB Score} & \textbf{0.68} \\
\bottomrule
\end{tabular}
\end{table}
To facilitate comparison between our agentic RAG-based system (\text{DR. INFO}) and general-purpose frontier LLMs, we include a visual benchmark alongside models such as o1, o3, GPT-4.1 Grok 3, Gemini 2.5 Pro, Claude 3.7 Sonnet, and Llama 4 Maverick. These models were evaluated on the same \text{HealthBench Hard} subset, and their axis-wise scores were obtained from published HealthBench results \cite{arora2024healthbench}.

\begin{figure}[h!]
    \centering
    \includegraphics[width=1.0\textwidth]{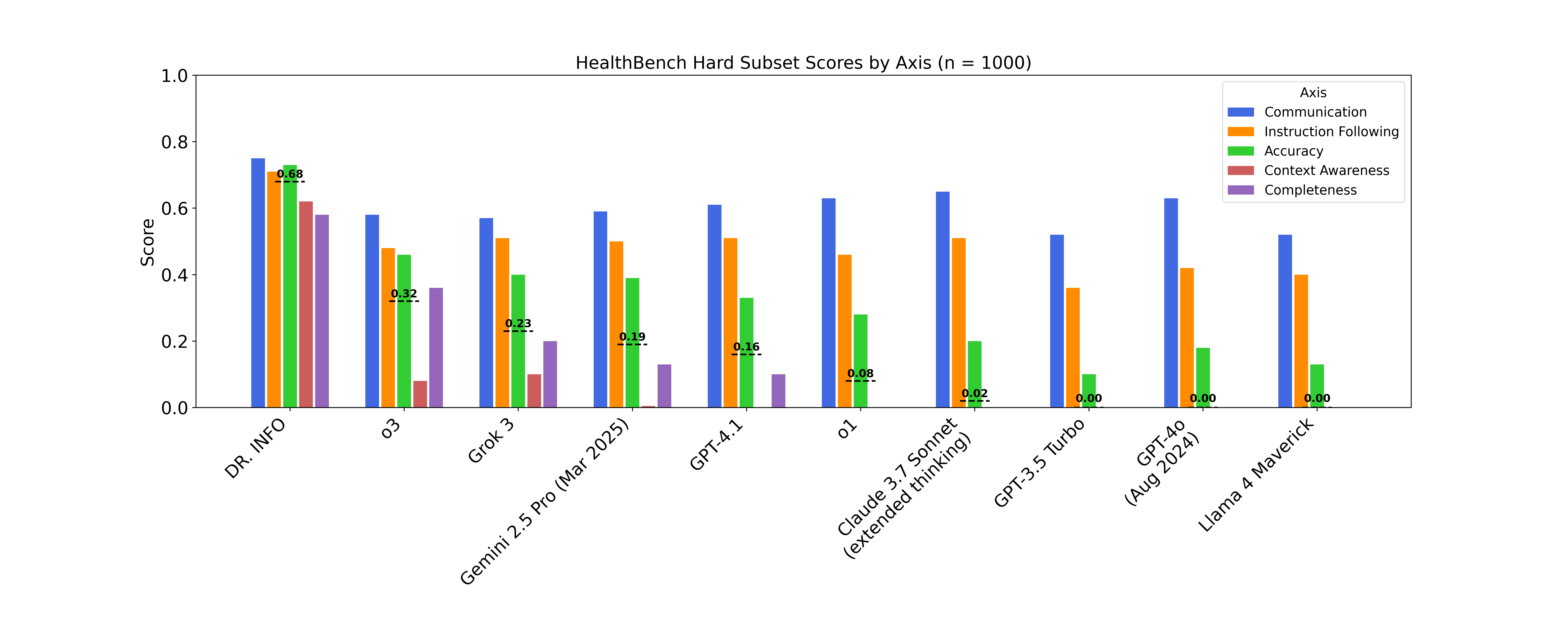} 
    \caption{Comparison of axis-wise scores for \text{DR. INFO} and other frontier LLMs on the HealthBench Hard subset. Model scores are visually approximated from \cite{arora2024healthbench}.}
    \label{fig:hard-results_frontier}
\end{figure}

OpenAI's GPT-5 model achieved a HealthBench score of \textbf{0.46} on the Hard subset, while GPT-5.2 (with thinking) reached \textbf{0.42} and GPT-5.1 (with thinking) achieved \textbf{0.40} \cite{openai2025gpt5} \cite{openai2025gpt52systemcard}. Our agentic RAG-based system, DR. INFO, achieved \textbf{0.68} on the HealthBench Hard subset across the evaluated axes, substantially outperforming the entire GPT-5 model family. As shown in Figure~\ref{fig:hard-results_frontier}, DR. INFO demonstrates superior performance across all five behavioral axes, with particularly notable advantages in context awareness and completeness compared to all frontier models. Axis-wise results for GPT-5 model family are unavailable, so it is omitted from Figure~\ref{fig:hard-results_frontier} but noted above for completeness.

\subsection{Comparison with Agentic/RAG based clinical assitants on Representative Subset 
(n=100)}

To further benchmark axis-level performance, we evaluated \text{DR. INFO} against other agentic RAG-based clinical assistants, namely OpenEvidence and Pathway (now DoxGPT, Doximity). We selected 100 randomly sampled, single-turn conversations from the HealthBench Hard dataset, ensuring even distribution across the seven thematic categories. All three systems were evaluated using HealthBench’s rubric framework on this representative subset.

\begin{figure}[h!]
    \centering
    \includegraphics[width=1.0\textwidth]{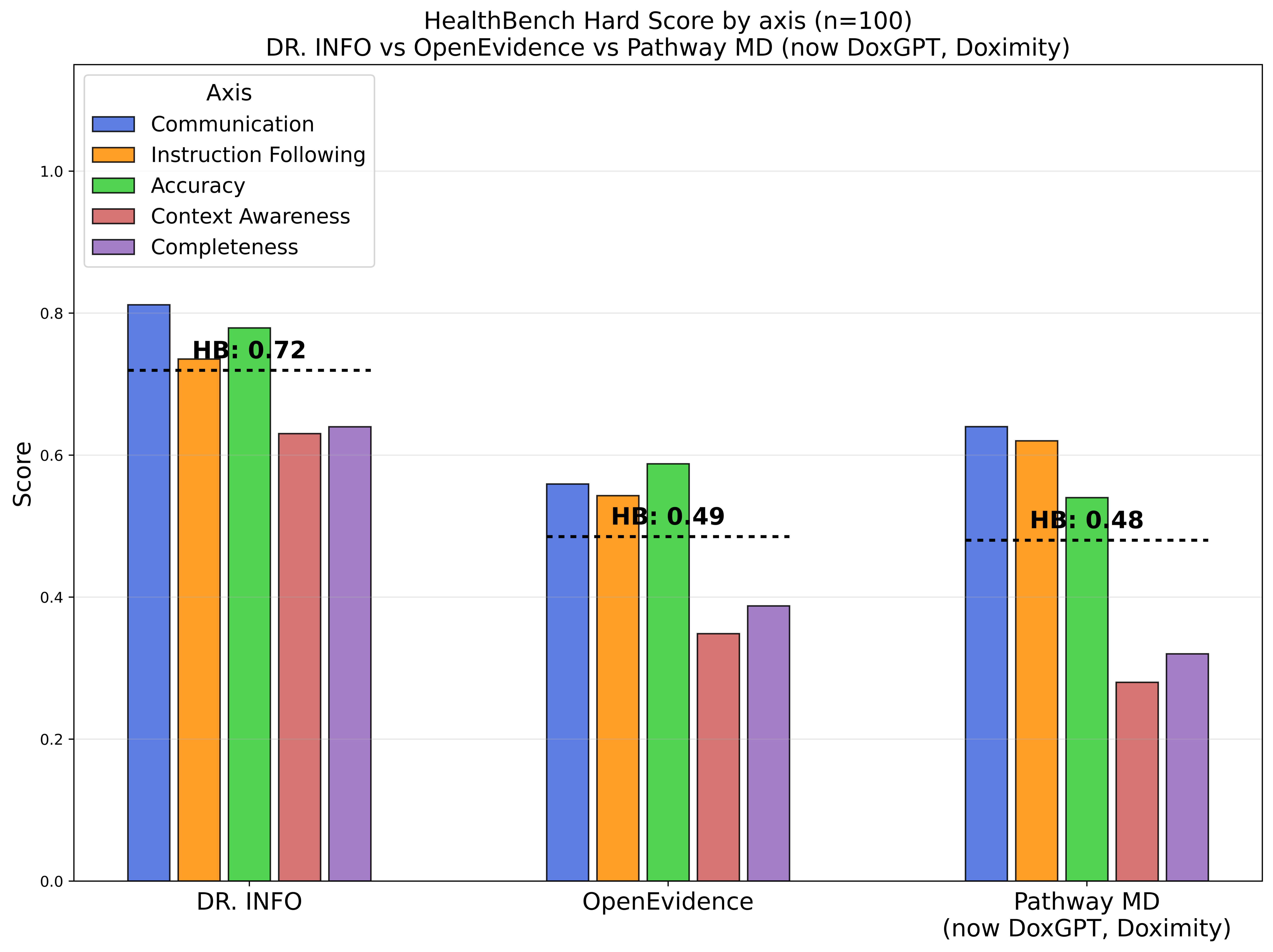} 
    \caption{HealthBench axis-wise scores for 100-sample subset: \text{DR. INFO} vs. OpenEvidence vs. Pathway.md (DoxGPT)}
    \label{fig:hb100-results}
\end{figure}

Figure~\ref{fig:hb100-results} shows the axis-level scores for each system. \text{DR. INFO} achieved a higher average HealthBench score of 0.72 on this evaluation set, compared to 0.49 for OpenEvidence and 0.48 for Pathway.\footnote{OpenEvidence was reevaluated in February 2026 with updated scoring; the scores remained consistent with the original evaluation.}

\begin{table}[h]
\centering
\caption{Axis-wise Scores on HealthBench Hard Subset (n=100)}
\label{tab:hb100}
\begin{tabular}{lccc}
\toprule
\textbf{Axis} & \textbf{\text{DR. INFO}} & \textbf{OpenEvidence} & \textbf{Pathway.md (DoxGPT)} \\
\midrule
Communication         & 0.81 & 0.56 & 0.64 \\
Instruction Following & 0.74 & 0.54 & 0.62 \\
Accuracy              & 0.78 & 0.59 & 0.54 \\
Context Awareness     & 0.63 & 0.35 & 0.28 \\
Completeness          & 0.64 & 0.39 & 0.29 \\
\midrule
\textbf{HB Score} & \textbf{0.72} & \textbf{0.49} & \textbf{0.48} \\
\bottomrule
\end{tabular}
\end{table}

\subsubsection{Comparative Performance with Bootstrapping (n=100)}
To assess the performance differences between the DR. INFO, OpenEvidence and Pathway.md (DoxGPT, Doximity), we conducted a statistical analysis on a representative subset of 100 samples from the HealthBench Hard set. The use of a smaller subset was necessitated by practical constraints, including rate limits and manual effort required for OpenEvidence and Pathway.md (now DoxGPT, Doximity).

To account for the limited sample size and the inherent stochasticity of LLM-based scoring, we employed a bootstrapping methodology. We performed 10,000 resampling iterations on the axis-level scores generated by the LLM rubric. In each iteration, these scores were sampled with replacement to compute an overall HealthBench score, allowing for the robust estimation of mean scores and confidence intervals (CIs). This analysis enabled a pairwise comparison to determine if the observed performance advantage of DR. INFO was statistically significant. The results are presented in Table~\ref{tab:hb_bootstrap_summary}.
\begin{table}[h!]
\centering
\caption{Bootstrapped HealthBench (HB) Scores and Significance Analysis ($n=100$)}
\label{tab:hb_bootstrap_summary}
\begin{tabular}{lccc}
\toprule
 & \textbf{HB Score} & \textbf{90\% CI} & \textbf{95\% CI} \\
\midrule
\text{DR. INFO}        & 0.72 & [0.6941, 0.7488] & [0.6890, 0.7542] \\
OpenEvidence   & 0.49 & [0.4545, 0.5206] & [0.4485, 0.5272] \\
Pathway.md (DoxGPT)    & 0.48 & [0.4415, 0.5220] & [0.4332, 0.5287] \\
\bottomrule
\end{tabular}

\vspace{1em}

\begin{tabular}{lcc}
\toprule
\textbf{Pairwise Comparison} & \textbf{CI} & \textbf{Significant?} \\
\midrule
\text{DR. INFO} $-$ OpenEvidence & [0.1908, 0.2768] & \textbf{Yes (90\%)} \\
\text{DR. INFO} $-$ Pathway.md (DoxGPT)   & [0.1929, 0.2903] & \textbf{Yes (90\%)} \\
\midrule
\text{DR. INFO} $-$ OpenEvidence & [0.1823, 0.2849] & \textbf{Yes (95\%)} \\
\text{DR. INFO} $-$ Pathway.md (DoxGPT)   & [0.1836, 0.2996] & \textbf{Yes (95\%)} \\
\bottomrule
\end{tabular}
\end{table}

\section{Discussion}

The evaluation of our agentic RAG-based clinical support assistant, \text{DR. INFO} \cite{drinfo_app}, using the HealthBench framework reveals valuable insights into its capabilities and limitations across real-world medical conversations. Through both large-scale evaluation on the HealthBench Hard subset ($n = 1000$) and a controlled comparison against similar systems on a smaller representative sample ($n = 100$), we observed consistent strengths in core clinical behaviors such as communication, instruction following, and factual accuracy.

On the 1000-sample HealthBench Hard subset, OpenAI's recently released GPT-5 model family \cite{openai2025gpt5, openai2025gpt52systemcard} represents the current state-of-the-art among frontier LLMs, with GPT-5 achieving 0.46, GPT-5.2 (thinking) achieving 0.42, and GPT-5.1 (thinking) achieving 0.40.\footnote{GPT-5 family axis-wise results are not publicly available; only overall HealthBench Hard scores are reported. These models are not included in Figure~\ref{fig:hard-results_frontier}.} In comparison, \text{DR. INFO} scored \textbf{0.68}, surpassing GPT-5 by 48\%, GPT-5.2 by 62\%, and GPT-5.1 by 70\%. Against other frontier models, DR. INFO outperformed Grok 3 by nearly 3$\times$ and Gemini 2.5 Pro by nearly 3.5$\times$. Notably, \text{DR. INFO} delivered approximately 4$\times$ higher performance in \textit{context awareness} (0.62 vs. 0.16 for Grok 3), where most frontier models exhibited little or no competence. Similarly, its \textit{completeness} score (0.58) exceeded Grok 3's (0.20) by nearly 3$\times$, reflecting stronger ability to generate responses that are thorough, well-reasoned, and supportable. Across other axes like \textit{communication} (0.75), \textit{instruction following} (0.71), and \textit{accuracy} (0.73), \text{DR. INFO} also consistently outperformed all evaluated frontier systems, indicating a higher degree of reliability and alignment in complex, high-stakes interactions.

To better understand performance relative to similar architectures, we conducted an additional 100-sample evaluation comparing \text{DR. INFO} with OpenEvidence and Pathway.md (now DoxGPT, Doximity), two other retrieval-augmented clinical support assistants. All these were evaluated using rubric-based scoring across a uniformly sampled set of single-turn conversations from the HealthBench Hard dataset. As shown in Table~\ref{tab:hb100} and Figure~\ref{fig:hb100-results}, \text{DR. INFO} achieved the highest average HealthBench score (0.72), ahead of OpenEvidence (0.49) and Pathway (0.48).

More specifically, \text{DR. INFO} demonstrated clear advantages across all axes. In \textit{communication} (0.81 vs. 0.56/0.64), \textit{instruction following} (0.74 vs. 0.54/0.62), \textit{accuracy} (0.78 vs. 0.59/0.54), \textit{context awareness} (0.63 vs. 0.35/0.28), and \textit{completeness} (0.64 vs. 0.39/0.29), DR. INFO substantially outperformed both OpenEvidence and Pathway.md (DoxGPT). These trends are illustrated in Figure~\ref{fig:hb100-results}, reinforcing the consistency and generalizability of our system across varied clinical scenarios and evaluation scales.

To quantify the robustness of these results given the limited sample size, we conducted a bootstrapped resampling analysis on the axis-level scores (n=100). The bootstrapped mean and confidence intervals (90\% and 95\%) for each system's HealthBench score are shown in Table~\ref{tab:hb_bootstrap_summary}. The differences in scores between \text{DR. INFO} and the baselines were statistically significant under both 90\% and 95\% CI thresholds. These results indicate that \text{DR. INFO} robustly outperforms the compared systems even under variability.

While this bootstrapped comparison was restricted to a 100-sample representative subset due to rate limits and manual evaluation constraints, we acknowledge that more definitive comparisons would require large-scale scoring across the full HealthBench Hard set. Expanding this analysis to the full 1000-sample dataset remains an important direction for future work, and would enable higher-confidence statistical conclusions about system-level differences.

While HealthBench provides a rigorous framework for evaluating behavioral competencies in clinical dialogue, it has limitations. Rubric-based assessments, though authored by physicians, may reflect subjective interpretations. Additionally, HealthBench is restricted to text-based interactions; it does not evaluate performance on other critical modalities such as medical imaging, genomics, structured data interpretation, or multimodal diagnostic tasks. A comprehensive assessment of clinical readiness would require benchmarking across these diverse modalities to reflect the full spectrum of real-world medical decision-making.

\section{Conclusion}
This article presents a comprehensive evaluation of \text{DR. INFO} \cite{drinfo_app}, an agentic retrieval-augmented clinical support assistant, using the HealthBench framework. Unlike conventional benchmarks such as the USMLE or MedQA, HealthBench enables a multi-dimensional evaluation of real-world clinical conversations across axes that matter for safe and effective deployment.

\text{DR. INFO} demonstrated consistent strengths across challenging scenarios achieving a HealthBench score of 0.68 on the Hard subset and outperforming previously reported baselines. Its high scores in communication, instruction-following, and factual accuracy suggest that retrieval-augmented, agentic strategies can support more aligned and clinically effective responses. Notably, the system maintained robust performance even in ambiguous and information-sparse scenarios, where conventional LLMs often fail. The rubric-based evaluation further enabled fine-grained behavioral analysis, surfacing areas for targeted refinement, particularly in context awareness and response completeness. These findings underscore the value of behavior-level, clinically grounded benchmarks like HealthBench for assessing real-world readiness in medical AI systems.

To validate the robustness of observed gains, we applied bootstrapped resampling on a representative subset of the dataset. The results showed that \text{DR. INFO} consistently outperformed comparable clinical assistants with statistical significance under both 90\% and 95\% confidence thresholds, providing strong evidence of performance advantages.

Together, these findings establish \text{DR. INFO} as a competitive, real-world-ready clinical assistant and reinforce the utility of behavior-level evaluation frameworks such as HealthBench in guiding the development and safe integration of LLM-based systems into healthcare. A HealthBench score of 0.68 on the Hard dataset represents state-of-the-art performance, outpacing the GPT-5 model family (GPT-5: 0.46, GPT-5.2: 0.42, GPT-5.1: 0.40) and all other frontier LLMs as well as similarly designed agentic systems. This underscores the broader challenge of clinical deployment and the importance of continued benchmarking and iteration.
\section{Acknowledgements}

We also extend our sincere gratitude to the \textbf{following physicians for their domain expertise and time, which were instrumental in enhancing \text{DR. INFO}'s clinical relevance and robustness across specialties}: Dr.~Rita Viana (Critical Care), Dr.~Fernando Correia (Internal Medicine), Dr.~Tiago Mendes (Internal Medicine), Dr.~Marta Isidoro (Anesthesiology), Dr.~Diogo Alves (Critical Care), Dr.~Luana Passos (Medical Student), Dr.~Joana Esteves (Critical Care), Dr.~Oriana Hierro (Family Medicine), Dr.~Diogo Caetano (Medical Student), Dr.~Francisco Silva (Critical Care), Dr.~Isa Agudo (Internal Medicine), Dr.~Maria João Araújo (Family Medicine), Dr.~Pedro Nunes (Pediatrics), Dr.~Ana Rita Branco (Critical Care), Dr.~Maria López (Family Medicine), Dr.~Diogo Nogueira (Medical Student), Dr.~Flávia Freitas (Internal Medicine), Dr.~José Diogo Martins (Internal Medicine), Dr.~Pedro Mesquita (Critical Care), Dr.~Maria Costa (Medical Student), Dr.~Eduardo Viana (Internal Medicine), Dr.~Marcos Pedro (Family Medicine), Dr.~Carla Costa Dias (Cardiology), Dr.~Hugo Daniel Silva (Anesthesiology), Dr.~Helder Novais e Bastos (Pulmonology), Dr.~André Colmente (Critical Care), Dr.~Lara Gonçalves (Medical Student), and Dr.~Carolina Araújo (Family Medicine).

We also gratefully acknowledge the technical support provided by Sonu Kumar, Luv Chafale, and Shubhayu Bhowmick.
\section{Conflicts of Interest}
During the drafting of the work, Sandhanakrishnan Ravichandran, Shivesh Kumar, Rog\'erio Corga Da Silva, Miguel Romano, Reinhard Berkels, Michiel van der Heijden, Olivier Fail, and Valentine Emmanuel reported employment with Synduct GmbH.
\section{Funding}
This work was supported by Synduct GmbH.

\printbibliography

\end{document}